\documentclass[superscriptaddress,floatfix,nofootinbib,longbibliography,prl,twocolumn]{revtex4-1}
\usepackage{graphicx,amsfonts,amssymb,amsmath,hyperref,hypcap,enumerate}
\usepackage{slashed}
\usepackage{amsmath,amssymb,bm,graphicx,dsfont}
\usepackage[latin9]{inputenc}
\usepackage{amsmath}
\usepackage{slashed}
\usepackage{dsfont}
\usepackage{amstext}
\usepackage{amssymb}
\usepackage{amsbsy}
\usepackage{amsthm}
\usepackage{epsfig}
\usepackage{graphicx}
\usepackage{bbm}
\usepackage{hyperref}
\usepackage{color}
\usepackage{booktabs}
\usepackage{array}
\usepackage{slashed}
\newcommand{\be}{\begin{equation}}
\newcommand{\ba}{\begin{align}}
\newcommand{\ee}{\end{equation}}
\newcommand{\bea}{\begin{eqnarray}}
\newcommand{\eea}{\end{eqnarray}}
\newcommand{\beq}{\begin{equation}}
\newcommand{\eeq}{\end{equation}}
\newcommand{\beqn}{\begin{eqnarray}}
\newcommand{\eeqn}{\end{eqnarray}}

\newcommand{\la}{\langle}
\newcommand{\ra}{\rangle}

\renewcommand{\vec}[1]{{\bf #1}}

\usepackage{soul}

\begin{document}
\title{\texorpdfstring{Theory of Dirac spin liquids on spin-$S$ triangular lattice: \\
		possible application to $\alpha$-CrOOH(D)}{DSL}}

\author{Vladimir Calvera}
\email{fvcalvera@stanford.edu}
\affiliation{Department of Physics, Stanford University, Stanford, CA 94305
}

\author{Chong Wang}%
\email{cwang4@pitp.ca}
\affiliation{%
	Perimeter Institute for Theoretical Physics, Waterloo, Ontario N2L 2Y5, Canada
}%
\date{\today}

\begin{abstract}
{
\noindent
Triangular lattice quantum antiferromagnet has recently emerged to be a promising playground for realizing Dirac spin liquids (DSLs) -- a class of highly entangled quantum phases hosting emergent gauge fields and gapless Dirac fermions. While previous theories and experiments focused mainly on $S=1/2$ spin systems, more recently signals of a DSL were detected in an $S=3/2$ system $\alpha$-CrOOH(D) in Ref.~\cite{Liuetal20}. 
In this work we develop a theory of DSLs on triangular lattice with spin-$S$ moments. 
We argue that in the most natural scenario, a spin-$S$ system realizes a $U(2S)$ DSL, described at low energy by gapless Dirac fermions coupled with an emergent $U(2S)$ gauge field (also known as $U(2S)$ QCD$_3$). An appealing feature of this scenario is that at sufficiently large $S$, the $U(2S)$ QCD becomes intrinsically unstable toward spontaneous symmetry breaking and confinement. The confined phase is simply the $120^{\circ}$ coplanar magnetic order, which agrees with semiclassical (large-$S$) results on simple Heisenberg-like models. Other scenarios are nevertheless possible, especially at small $S$ when quantum fluctuations are strong. For $S=3/2$, we argue that a $U(1)$ DSL is also theoretically possible and phenomenologically compatible with existing measurements. 
One way to distinguish the $U(3)$ DSL from the $U(1)$ DSL is to break time-reversal symmetry, for example by adding a spin chirality term $\vec{S}_i\cdot(\vec{S}_j\times\vec{S}_k)$ in numerical simulations: the $U(1)$ DSL becomes the standard Kalmeyer-Laughlin chiral spin liquid with semion/anti-semion excitation; the $U(3)$ DSL, in contrast, becomes a non-abelian chiral spin liquid described by the $SU(2)_3$ topological order, with Fibonacci-like anyons. 


 }
\end{abstract}

\maketitle


Triangular lattice antiferromagnet was historically the first system suggested, by Anderson\cite{Anderson1973}, to realize quantum spin liquid states\cite{savary_2017,Zhou_QSL_review}. Although the ground state of the nearest-neighbor Heisenberg model forms a classical $120^{\circ}$ coplanar magnetic order even for $S=1/2$\cite{120order1,120order2,120order3,120order4}, the order can be destroyed upon including some relatively weak further-neighbor couplings. For example, for $S=1/2$ systems with a second-neighbor exchange $J_2$, multiple numerical studies indicate the existence of a quantum spin liquid phase  in the range $0.07 \leq J_2/J_1\leq 0.15$, sandwiched between the familiar $120^{\circ}$ coplanar magnetic order at $J2/J1< 0.07$ and a strip magnetic order at $J_2/J_1> 0.15$\cite{J1J2_1,J1J2_2,J1J2_3,J1J2_4,J1J2_5,J1J2_6,J1J2_7,J1J2_8,J1J2_9,J1J2_10,J1J2_11,Ferrari19,He19} . 

It has become increasingly clear recently that the quantum spin liquid state realized this way, at least on $S=1/2$ triangular lattice, is likely a $U(1)$ Dirac spin liquid ($U(1)$ DSL). This conclusion comes from a synthesis of significant progresses in formal theories, numerical simulations and experimental observations: the stability of DSL as a gapless phase was established through a series of theoretical arguments\cite{kapustin_2002,hermele_20041,song19a,Songetal} and lattice simulations\cite{KarthikQED,KarthikQEDa,KarthikMonopole}. It was also shown theoretically that the spin-spin correlation function in the DSL has nontrivial weight at the Brillouin zone corners (the $\vec{K}=(2\pi/3,-2\pi/3)$ and $\vec{K}'=-\vec{K}$ points), which comes from monopole fluctuations in the $U(1)$ gauge field and cannot be understood from spinon mean field theory\cite{song19a,Songetal}. This spectral weight was observed in numerical simulations of the $J_1-J_2$ Heisenberg model\cite{Ferrari19,He19}, as well as in neutron scattering experiments on powders of NaYbO$_2$, an $S=1/2$ triangular system\cite{NaYbO2Neutron}. Although more refined experimental confirmations are still needed, there are good reasons to be optimistic on this front. 

More recently\cite{Liuetal20} DSL-like behaviors were also observed in an $S=3/2$ triangular system $\alpha$-CrOOH(D) (delafossites green-grey powder): the system does not order down to $\sim 2K$ which is much lower than the Curie temperature $\theta_{CW}\sim-211.6(5)K$;  the specific heat behaves roughly as $C\sim T^{2.2}$ which is consistent with a relativistic critical state with $C\sim T^2$; furthermore the spectral weight in neutron scattering (again on powders) accumulates around the $\vec{K}, \vec{K}'$ points which is similar to the $S=1/2$ case. Assuming these signatures indeed come from a spin liquid phase at low energy, the obvious question is: what kind of spin liquid is it? Is it the same spin liquid as the $S=1/2$ $U(1)$ DSL in terms of universal properties, or is it a distinct phase?

We can even consider a more general setting. Assuming that a $U(1)$ DSL is indeed realized, say in the $S=1/2$  $J_1-J_2$ model in appropriate regimes, one can ask how this spin liquid evolves as we increase $S$ (with the form of the Hamiltonian fixed). For sufficiently large spin (beyond some critical value $S_c$) the system becomes semi-classical and the spin liquid should disappear from the phase diagram. Indeed for very large $S$ where $1/S$-expansion becomes reliable, it is known that for small $J_2/J_1$ the system forms the well known $120^{\circ}$ coplanar anti-ferromagnet, and when $J_2/J_1$ exceeds a critical value $\sim 1/8$ the system goes through a first order transition and forms a stripe antiferromagnet\cite{Stripe1, Stripe2}. The experiments on $\alpha$-CrOOH(D)  suggests that $S_c>3/2$, so the natural question is: what types of spin liquids should we expect for $1/2<S<S_c$?

In order to develop a theory of critical Dirac spin liquids for general $S$, we start with a parton decomposition of the spin operators (for a general review of the parton approach see Ref.~\cite{wenbook}):
\be
\vec{S}_{i,\mu}=\frac{1}{2}\sum_{a=1}^{2S}\sum_{\alpha,\beta=\uparrow,\downarrow}f^{\dagger}_{i,a,\alpha}{\sigma}_{\mu}^{\alpha\beta}f_{i,a,\beta},
\ee
where $i$ is the lattice site index, $\mu=x,y,z$ denotes the spin components, $\sigma_{\mu}$ is the Pauli matrix and $f_{i,a,\alpha}$ is a fermion annihilation operator that carries spin-$1/2$ (indexed by $\alpha$) and a color index $a=1,2...2S$. The auxiliary fermion $f$ is also known as the spinon. This representation of the spin operator introduces some redundancies. The physical spin Hilbert space can be recovered by imposing the constraints of single occupancy for each color $a$ ($\sum_{\alpha}f^{\dagger}_{\alpha,a}f_{\alpha,a}=1$) and total symmetrization over all different colors. These constraints can be conveniently formulated as a local $Sp(2S)$ gauge invariance on the spinons: 
\be
\left(\begin{array}{c}
    f_{1,\uparrow}  \\
    \vdots \\
    f_{2S,\uparrow}  \\
    f^{\dagger}_{1,\downarrow}  \\
    \vdots  \\
    f^{\dagger}_{2S,\downarrow}  \\
\end{array} \right) \longrightarrow U \left(\begin{array}{c}
    f_{1,\uparrow}  \\
    \vdots \\
    f_{2S,\uparrow}  \\
    f^{\dagger}_{1,\downarrow}  \\
    \vdots  \\
    f^{\dagger}_{2S,\downarrow}  \\
\end{array} \right),
\ee
where the $Sp(2S)$ matrix $U$ can vary from site to site. For $S=1/2$, this recovers the more familiar $Sp(1)=SU(2)$ gauge symmetry.

We should now proceed with a spinon mean field ansatz. Motivated by the anzatz for the $U(1)$ DSL in $S=1/2$ systems, we use the following mean field Hamiltonian:
\be
\label{eq: TDiracansatz}
H_{MF}^{S}=-t\sum_{a=1}^{2S}\sum_{<ij>}(-1)^{h_{ij}}(f_{i,a,\alpha}^{\dagger}f_{j,a,\alpha}+h.c.)
\ee
where the sign factor $(-1)^{h_{ij}}$ gives a $\pi$-flux on all upward triangles and zero flux on all downward triangles. For later convenience, we assume $t>0$ without loss of generality. This ansatz breaks the microscopic $Sp(2S)$ gauge symmetry down to $U(2S)$:
\be
f_{a,\alpha}\to U_{ab}f_{b,\alpha},
\ee
where $U\in U(2S)$. The mean field ansatz also satisfies the average constraint $\la \sum_{\alpha}f^{\dagger}_{a,\alpha}f_{a,\alpha}\ra=1$ for each color $a$. The hopping amplitude is designed so that at low energy the spinons form gapless Dirac cones at two points in the reduced Brillouin zone (two valleys). Including the degeneracy from the spin and color indices, there are in total $4\times 2S$ two-component Dirac fermions at low energy, and they couple to a $U(2S)= (U(1)\times SU(2S))/\mathbb{Z}_{2S}$ gauge field in the fundamental representation. The resulting low energy theory is a $U(2S)$ gauge theory with $N_f=4$ flavors of fundamental Dirac fermions. This ansatz is very similar to that of the standard $U(1)$ DSL for $S=1/2$ -- the only difference is that we have $2S$ colors instead of only $1$ color for $S=1/2$, and we shall call this state a $U(2S)$ DSL. The low energy theory, known as $U(2S)$ QCD$_3$ with $N_f=4$, is described by the following Lagrangian:
\be
\label{eq: QCDL}
\mathcal{L}^{(S)}=\sum_{i=1}^{4}\bar{\psi}_ii\slashed{D}_{\boldsymbol{\mathcal{A}}}\psi_i+...,
\ee
where $\boldsymbol{\mathcal{A}}$ is a $U(2S)$ gauge field and $\psi_i$ is a Dirac fermion that transforms as a $U(2S)$ fundamental and a two-component spinor under the emergent Lorentz symmetry. The $...$ represents additional terms allowed by physical symmetries such as four-fermion interactions and monopole tunnelings. For $S=1/2$ we recover the QED$_3$ description of the standard $U(1)$ DSL.

Now what is the fate of the theory Eq.~\eqref{eq: QCDL} at low energy? It is known that when $N_f \gg 2S$, the Dirac fermions and gauge fields all remain gapless at low energy and the theory flows under renormalization group (RG) to a conformal field theory (CFT). An emergent $SU(N_f)=SU(4)$ symmetry is expected among the Dirac fermions ($2$ spins and $2$ valleys) in the CFT\cite{hermele_2005_mother}. For $N_f=4$, the conformality remains as long as $2S<N_c$ for some $N_c$. The exact value of $N_c$ is currently unknown and requires future numerical simulation. Based on the apparent stability of $U(1)$ DSL we expect $N_c>1$. If $N_c>3$, the $U(3)$ DSL will be a stable gapless phase that could be realized in materials like $\alpha$-CrOOH(D).\footnote{To be more precise: the stability of $U(2S)$ DSL requires not only the conformality of the pristine $U(2S)$ QCD$_3$, but also the irrelevance of operators that are allowed by microscopic symmetries (the $...$ terms in Eq.~\eqref{eq: QCDL}). For the purpose of this paper we do not carefully distinguish the two types of instabilities. However this will be important in future numerical studies to decide the critical value $S_c$.}

In $U(1)$ DSL for $S=1/2$ systems, monopole operators of the $U(1)$ gauge field form an important class of critical fluctuation\cite{kapustin_2002,hermele_20041,DyerMonopoleTaxonomy,song19a,Songetal}. This is also true for general $U(2S)$ DSLs. To introduce the monopoles in $U(2S)$ QCD$_3$, let us first write the $U(2S)$ gauge field as
\be
\boldsymbol{\mathcal{A}}=\mathcal{A}+\frac{1}{2S}aI,
\ee
where $\mathcal{A}$ is an $SU(2S)$ gauge field while $a$ is a $U(1)$ gauge field ($I$ being the $2S\times 2S$ identity matrix). The normalization factor $1/2S$ above is chosen so that the minimum $U(1)$ gauge charge carried by $SU(2S)$-invariant operators is unity. We can now define monopoles for the $a$ gauge field as operators that insert nontrivial $U(1)$ gauge flux $\int da$. The Dirac quantization rule states that the monopole should be local with respect to operators that are gauge-invariant under other gauge symmetries (here being the $SU(2S)$). Therefore the minimum $U(1)$ monopole carries $2\pi$ flux of $a$. Equivalently, the conserved flux current is
\be
j^{\mu}_{\rm{flux}}=\frac{1}{2\pi}\epsilon^{\mu\nu\lambda}\partial_{\nu}a_{\lambda}.
\ee
One subtlety compared to $U(1)$ DSL is that the monopoles are not directly local with respect to the Dirac fermions $\psi$, since they couple to $a$ as charge $1/2S$ fields. This means that if the $U(2S)$ QCD remains gapless at low energy and the $\psi$ fermions are deconfined, the $U(1)$ monopole must be accompanied by an $SU(2S)$ flux, so that it can be local with respect to $\psi$. Here it is useful to consider a $U(1)^{\otimes 2S}$ subgroup of $U(2S)$ and denote the gauge field for each $U(1)$ subgroup as $\mathcal{A}^{(a)}$, namely $\boldsymbol{\mathcal{A}} =\textit{diag}(\mathcal{A}^{(a)})$. We can now label the flux configuration by specifying the gauge flux of each $\mathcal{A}^{(a)}$: 
\be
\int d\mathcal{A}^{(a)}=2\pi n_a, \hspace{5pt}n_a\in\mathbb{Z}.
\ee
The topological flux of the $a$ gauge field is 
\be
\int da=2\pi\sum_an_a,
\ee
and the most relevant monopole (the one with the lowest scaling dimension or energy) corresponds to 
\be
(n_1,n_2...n_{2S})=(1,0,...0),
\ee
or other configurations obtained from it through $SU(2S)$ rotations. This has been quantitatively discussed using a large $N_f$ expansion in Ref.~\cite{DyerMonopoleTaxonomy}. For $N_f\gg 2S$, the scaling dimension of the fundamental monopole is
\be
\label{eq: largeNDelta}
\Delta_{\mathcal{M}}=0.265N_f-0.0383-0.516(2S-1)+O(1/N_f)
\ee
Although we only have $N_f=4$, the qualitative aspects (such as symmetry quantum numbers) of the most monopole operators are expected to remain the same as long as the theory stays gapless. 


The above minimum monopole, denoted as $\mathcal{M}_{\mathcal{A}}$, effectively behaves as a $U(1)$ monopole seen only by one color of Dirac fermion $\psi_{a=1}$. Since $\psi_{a=1}$ has exactly the same band structure as the Dirac fermions in the $U(1)$ DSL, many properties of $\mathcal{M}_{\mathcal{A}}$ will be identical to the monopoles in the $U(1)$ DSL. Each two-component Dirac fermion with color index $a=1$ contributes a zero mode in the flux background $\chi_i$ ($i=1,...4$). Gauge invariance requires filling half of the zero modes, so that a gauge-invariant monopole operator looks like\cite{kapustinqed}:
\be
\chi^{\dagger}_i\chi^{\dagger}_j\mathcal{M}_{\mathcal{A}},
\ee
where $\mathcal{M}_{\mathcal{A}}$ is a ``bare'' flux-insertion operator with all the zero modes empty. There are in total $C^{4}_2=6$ such operators that form a vector representation of the $SO(6)=SU(4)/\mathbb{Z}_2$ flavor symmetry. Three of these six operators form a triplet under the microscopic spin $SO(3)$ and the remaining three are singlets. The spin triplet monopole makes the most important contribution to the spectral weight of spin-spin correlation function. These monopoles could also carry nontrivial lattice momenta, and the neutron spectral weight will accumulate at the momentum of the spin triplet monopole. The monopole momentum comes from a nontrivial Berry phase as the gauge flux moves in a lattice of gauge charges, and the pattern of the gauge charges is determined by the topology of the underlying spinon band structure. It was shown in Refs.~\cite{song19a, Songetal} that for the triangular mean field ansatz Eq.~\eqref{eq: TDiracansatz}, the spin triplet monopoles carry a lattice momentum $\vec{K}=(2\pi/3,-2\pi/3)$. This $\vec{K}$ momentum makes the monopole sharply distinct from spinon scattering operators $\psi^{\dagger}_i\psi_j$, which can at most have a lattice momenta at the $\vec{M}$ points ($(\pi,0)$, $(0,\pi)$ and $(\pi,\pi)$) according to their band structure. This is why observations of spectral weight at $\vec{K}$, $\vec{M}$ and symmetry related momenta are considered important evidences for the existence of the DSL. The same analysis shows that the three spin-singlet monopoles carry momenta $\vec{K}+\vec{M}$ -- this will be important for measurements on singlet excitations, such as X-ray scatterings.


We now ask what happens if $S>S_c$ and the $U(2S)$ QCD$_3$ becomes unstable. The leading instability of $U(2S)$ QCD is believed to be a spontaneous breaking of the  $SU(N_f)$ flavor symmetry down to $SU(N_f/2)\times SU(N_f/2)\times U(1)$ (recall that $N_f$ is even due to fermion doubling from parity anomaly)\cite{QCD3,QCD3symmbreak}. One can think of this ``spontaneous chiral symmetry breaking'' as the formation of a Dirac mass term of the form, up to $SU(N_f)$ flavor rotations,
\be
\label{eq: chiralmass}
m\bar{\psi}\sigma^z\psi,
\ee
which has value $+m$ for $N_f/2$ flavors of Dirac fermions and $-m$ for the other $N_f/2$ flavors. Below this mass scale only the $U(2S)\sim SU(2S)\times U(1)$ gauge field remains. The $SU(2S)$ gauge field is expected to eventually become gapped and confine the $\psi$ fermions at lower energy, and only $SU(2S)$-invariant objects such as $\psi_{a=1}\psi_{a=2}...\psi_{a=2S}$ remain as gapped excitations. The $U(1)$ gauge field now remains as a free photon field in the IR.\footnote{In fact this sequence of events should be viewed as a picture to aid our thinking. In general they will not be clearly separated in energy scale, unless $S$ happens to be barely above $S_c$.  } It is well known that a free $U(1)$ gauge field in $(2+1)d$ should be viewed as a conventional symmetry breaking phase, a ``superfluid'' that breaks the $U(1)$ flux conservation symmetry, with the photons being the Goldstone bosons. The free $U(1)$ gauge theory has a unique monopole operator, which serves as the condensed order parameter of this superfluid.

Now back to the $U(2S)$ DSLs on triangular lattice, with fixed $N_f=4$. For $S>S_c$, we expect the spontaneous chiral symmetry breaking $SU(4)\to SU(2)\times SU(2)\times U(1)$ followed by the confinement of the $SU(2S)$ gauge field. Microscopically we do not have the full $SU(4)$ symmetry, so the ordering pattern will be decided by microscopic details and cannot be determined from the effective field theory. For reasons that will become clear later, we consider a symmetry breaking pattern represented by the mass Eq.~\eqref{eq: chiralmass}, where $\sigma^z$  refers to a Pauli matrix in the spin index. This mass term produces a quantum spin Hall insulator with a mutual Chern-Simons term
\be
\frac{1}{2\pi}A^{S_z}da,
\ee
where $A^{S_z}$ is a probe $U(1)$ gauge field that couples to the conserved $S_z$. This term assigns a unit $S_z$ spin to each flux quanta of $a$, so that a monopole operator $\mathcal{M}$ behaves like $S_x+iS_y$. This monopole is a linear combination of the original $SO(6)$-vector monopoles in the QCD$_3$ theory -- the 6-fold degeneracy is now lifted due to the chiral symmetry breaking. In particular, the $\mathcal{M}$ operator inherits the lattice momentum of the spin-triplet monopoles of the QCD$_3$ theory. So $\mathcal{M}$ should carry lattice momentum $\vec{K}$. An operator $S_x+iS_y$ with lattice momentum $\vec{K}$ is nothing but the order parameter of the $120^{\circ}$ magnetic order, so the resulting state from spontaneous chiral symmetry breaking and confinement is simply the familiar $120^{\circ}$ order. Notice that when $S<S_c$, this instability towards $120^{\circ}$ order does not happen spontaneously, but can nevertheless take place as we drive the $U(2S)$ DSL through a (likely continuous) phase transition, which is described by a QCD-Gross-Neveu field theory.

The above analysis motivates us to conjecture that a $U(2S)$ DSL is realized in the  spin-$S$ $J_1-J_2$ model on triangular lattice, at some intermediate $J_2/J_1$. The appealing feature of this conjectured scenario is that it naturally reproduces the following nontrivial facts about the $J_1-J_2$ model:
\begin{enumerate}
\item At small $J_2/J_1$ the ground state forms a simple $120^{\circ}$ coplanar order for any $S$.
\item At $S=1/2$ the model appears to realize a $U(1)$ DSL phase at intermediate $J_2/J_1$.
\item At large $S$ the model becomes semiclassical with no spin liquid in the phase diagram.
\end{enumerate}

 \begin{figure}
 \begin{center}
\includegraphics[width=0.5\textwidth]{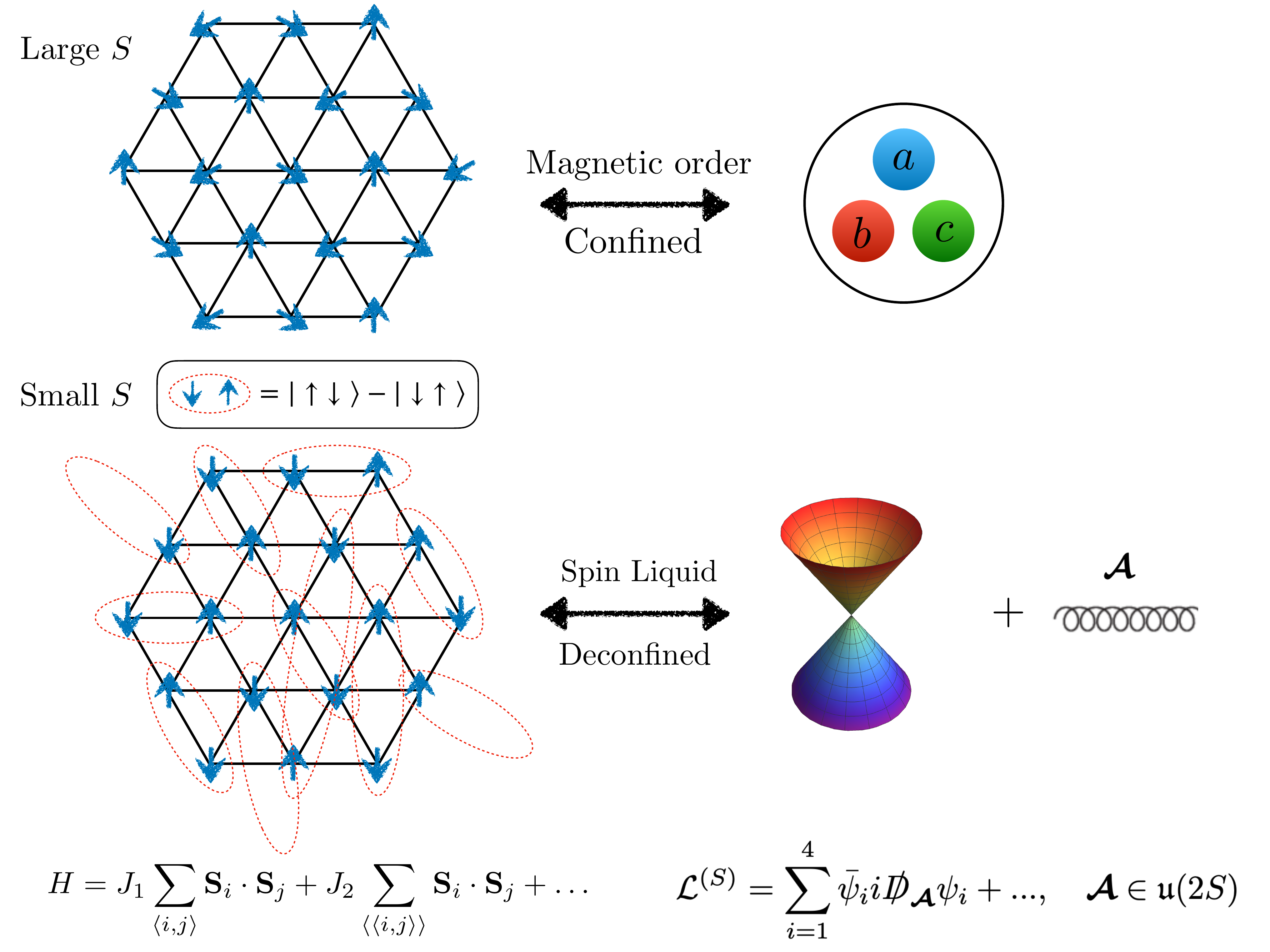}
 \end{center}
 \caption{We conjecture that the two theories in this figure are dual to each other in the IR. Left: Spin $S$ triangular lattice antiferromagnet with nearest and next-nearest Heisenberg couplings $J_1$, $J_2$, roughly in the regime $J_2/J_1\sim 0.1$. Right: $U(2S)$ QCD$_3$ with $N_f=4$. This is a strong-weak duality in the following sense: at large $S$ the triangular lattice antiferromagnet is weakly fluctuating and forms a classical order, while the $U(2S)$ QCD is strongly fluctuating and confines at low energy; at small $S$ (say $S=1/2$) the quantum fluctuations of the lattice spins lead to a spin liquid phase with resonating (fluctuating) singlet configurations, while the QCD becomes stable at low energy and stays gapless and deconfined.
 } \label{StrongWeak}
 \end{figure}

Our conjecture can also be phrased as an intriguing strong-weak duality between triangular lattice antiferromagnets and $U(N)$ QCD$_3$ with $N_f=4$, as we illustrate in Fig.~\ref{StrongWeak}. 

The above non-abelian scenario, although natural, is not the only possible scenario, especially at small $S$ (like $S=1$ or $3/2$). It is possible that the omitted $...$ terms in Eq.~\eqref{eq: QCDL} are strong enough to drive the system to a different phase, for example by spontaneously breaking (Higgsing) the $U(2S)$ gauge symmetry to a subgroup and gapping out some of the Dirac fermions. Such ``descendent'' states can be described in the parton approach by adding additional terms in the mean field Hamiltonian Eq.~\eqref{eq: TDiracansatz}. In general a $U(2S)$ DSL can have many different possible descendent phases. In the Appendix we discuss a number of such phases for $S=3/2$. Motivated by experiments on  $\alpha$-CrOOH(D) for $S=3/2$, we can further demand the descendent phase to be (1) gapless with critical spin fluctuations at $\vec{K}$ momentum and (2) in proximity to the $120^{\circ}$ order through a continuous phase transition. It turns out that  there is a unique descendent phase which satisfy these demands: a $U(1)$ DSL with the same universal properties as realized in $S=1/2$ systems. The corresponding parton mean field Hamiltonian is the original Eq.~\eqref{eq: TDiracansatz} plus the following perturbation:
\be
\label{eq: latticegap}
\delta H=\Delta\sum_{\la\la ij\ra \ra}(-1)^{s_{ij}}(\epsilon_{\alpha\beta}f^{\dagger}_{i,a=2,\alpha}f^{\dagger}_{j,a=3,\beta} +h.c.).
\ee
The above term is a second-neighbor $p$-wave pairing between the $a=2,3$ colors, where the factor $(-1)^{s_{ij}}$ inverts sign when $i,j$ are inverted, and the direction is chosen so that the flux on each triangle formed by one second neighbor and two nearest neighbors is uniformly $1$, namely $(-1)^{h_{ij}+h_{jk}+s_{ik}}=1$ when $i,j,k$ runs counterclockwise on the triangle. This second-neighbor term gaps out the Dirac cones from the two colors $a=2,3$ without breaking the physical symmetries, and breaks the gauge symmetry from $U(3)$ down to $U(1)\times U(1)$. The first $U(1)$ gauge symmetry acts as $f_{a=1}\to e^{i\theta}f_{a=1}$, while the second $U(1)$ gauge symmetry acts as $(f_{a=2},f_{a=3}^{\dagger})\to e^{i\theta'}(f_{a=2},f_{a=3}^{\dagger})$. The gauge field from the second $U(1)$ gauge symmetry couples only to $a=2,3$ fermions which are now gapped, so at low energy it should confine due to monopole proliferation. In the Appendix we show that the proliferated monopole here does not break any physical symmetry. This leaves behind a single $U(1)$ gauge field coupled with the Dirac fermions from $f_{a=1}$, which is nothing but a $U(1)$ DSL. The band structure for $f_{a=1}$ is identical to the spinons in the standard $U(1)$ DSL for $S=1/2$ systems, so we expect the same universal properties. In particular, the most important spin fluctuations come from the monopoles at $\vec{K}$ momentum, and the most important singlet fluctuations come from monopoles at $\vec{K}+\vec{M}$.

The above procedure using terms like Eq.~\eqref{eq: latticegap} can also be applied to other $S>1/2$. For any half-integer $S$ the simplest descendent state is the $U(1)$ DSL, while for integer $S$ the simplest descendent is a symmetric, gapped state with short-range entanglement. This scenario is similar to the Heisenberg spin chains in one dimension.

The two plausible DSLs for $S=3/2$, one with $U(3)$ gauge symmetry and the other with $U(1)$, are both compatible with existing measurements. They will differ in more refined details including various critical exponents, but it is quite nontrivial to either compute or measure such quantities. A natural question is whether we can distinguish the two scenarios using some readily accessible tool. We now show that a promising way to distinguish the different scenarios is to break time-reversal symmetry and measure the resulting topological order, which is achievable in numerical calculations, for example using the density-matrix-renormalization-group (DMRG) approach.

Specifically, we can introduce  a spin chirality term $\vec{S}_i\cdot(\vec{S}_j\times \vec{S}_k)$ which breaks time-reversal and reflection but preserve all other physical symmetries. For both the $U(3)$ and $U(1)$ DSLs, this spin chirality term will induce an $SU(4)$-symmetric, gauge-invariant mass gap for the Dirac fermions $\bar{\psi}\psi$. The gapped Dirac fermions form nontrivial Chern bands and generate a Chern-Simons (CS) term for the $U(3)$ or $U(1)$ gauge field. The resulting state is therefore a gapped topological order. For $U(1)$ DSL this is the well known Kalmeyer-Laughlin chiral spin liquid with semion/anti-semion excitations\cite{KalmeyerLaughlin}.   For $U(3)$ DSL the gapped Dirac fermions generate a CS term for the $U(3)$ gauge field $\mathcal{A}$:
\be
\pm\frac{2}{4\pi}{\rm{Tr}}\left(\mathcal{A}d\mathcal{A}-\frac{2i}{3}\mathcal{A}^3\right),
\ee
which is also known as $U(3)_{\pm2}$ CS theory. By level-rank duality, this theory with fermionic charge is dual to the $SU(2)_{\mp3}$ CS theory with bosonic charge. This topological order is non-abelian, with Fibonacci-like anyons. The two topological orders can be readily distinguished numerically, for example by examining the entanglement spectrum.

To summarize, we have discussed two possible gapless Dirac spin liquids -- with gauge symmetries $U(3)$ and $U(1)$, respectively -- that are compatible with recent experiments on the spin-$3/2$ triangular lattice system $\alpha$-CrOOH(D). The $U(3)$ DSL appears to be theoretically more natural but the $U(1)$ DSL is also possible. The two states can be distinguished, at least numerically, by examining the resulting topological order upon time-reversal breaking. Our discussion also calls for careful numerical studies of the $U(N)$ QCD$_3$ theory, for example using lattice gauge theory simulations to determine the critical value $N_c$. 

{\textbf{Acknowledgements}}: We thank Yin-Chen He, Tim Hsieh, Sung-Sik Lee and Liujun Zou for helpful discussions. Research at Perimeter Institute is supported by the Government of Canada through the Department of Innovation, Science and Economic Development Canada and by the Province of Ontario through the Ministry of Research, Innovation and Science.



\bibliography{SDSL}


\appendix

\section{Alternative DSLs on \texorpdfstring{$S=3/2$}{S=3/2} triangular lattice}
\label{appendix:alternative}

We start with the parton decomposition used in the main text for the $S=3/2$ operators:
\be
\vec{S}_{i,\mu}=\frac{1}{2}\sum_{a=1}^3\sum_{\alpha,\beta=\uparrow,\downarrow}f^{\dagger}_{i,a,\alpha}{\sigma}_{\mu}^{\alpha\beta}f_{i,a,\beta},
\ee
which comes with a microscopic $Sp(3)$ gauge symmetry. Below we describe a sequence of mean field ansatz, all with gapless Dirac fermions at low energy, starting from the simplest $U(3)$ Dirac ansatz used in the main text:
\begin{enumerate}
\item $U(3)$: consider nearest-neighbour hopping Hamiltonian with a staggered $\pi$-flux
\be
\label{U(3)ansatz}
H_{1}=-t\sum_{ij}(-1)^{h_{ij}}f^{\dagger}_{i,a,\alpha}f_{j,a,\alpha},
\ee
where the sign factor $(-1)^{h_{ij}}$ gives a $\pi$-flux on all upward triangles and zero flux on all downward triangles.  This ansatz leads to the $U(3)$ Dirac spin liquid discussed in the main text.


\item $U(2)\times \mathbb{Z}_2$: starting from the $U(3)$ theory, we can introduce a pairing term for a single color (say $a=3$) in the mean field theory to break the $U(3)$ gauge symmetry down to $U(2)\times\mathbb{Z}_2$:
\be
H_3=\sum_i\Delta f_{i,a=3,\uparrow}f_{i,a=3,\downarrow}+h.c.,
\ee
which also gaps out the $f_{a=3}$ Dirac fermions. This leaves behind a $U(2)$ gauge theory with $N_f=4$ at low energy, together with a gapped sector with $\mathbb{Z}_2$ topological order. As discussed in Refs~\cite{ZhengMeiQi15,Lu15} there are several different versions of $\mathbb{Z}_2$ topological orders depending on the exact form of pairing (in modern language they are topological orders enriched by spin and lattice symmetries in different ways). However as the gapped sector do not affect universal behaviours of physical correlation functions, we will not distinguish different (symmetry-enriched) topological orders here.

\item $U(1)\times \mathbb{Z}_2\times \mathbb{Z}_2$: starting from the $U(2)\times \mathbb{Z}_2$ state, we can further pair and gap out the $a=2$ fermions. This leaves behind a QED$_3$ with two gapped $Z_2$ topological ordered sectors.

\item $U(1)_A$: we now start from the $U(3)$ ansatz, and add the following term on lattice: 
\bea\label{HamA}
H_4=&& t'\sum_{\la \la ij\ra\ra}(-1)^{s_{ij}}(f^{\dagger}_{i,a=2,\alpha}f_{j,a=3,\alpha} +h.c.) \nonumber \\
&&+h\sum_{i}(f^{\dagger}_{i,a=1,\alpha}f_{i,a=2,\alpha}+h.c.).
\eea
The first term is a second-neighbor $p$-wave hybridyzation between the $a=2,3$ colors, where the factor $(-1)^{s_{ij}}$ inverts sign when $i,j$ are inverted, and the direction is chosen so that the flux on each triangle formed by one second neighbor and two nearest neighbors is uniformly $1$, namely $(-1)^{h_{ij}+h_{jk}+s_{ik}}=1$ when $i,j,k$ runs counterclockwise on the triangle. This second-neighbor term gaps out the Dirac cones formed by the two colors $a=2,3$, and breaks the gauge symmetry from $U(3)$ down to $U(1)\times U(1)\times U(1)$, but breaks no physical symmetry. The second on-site hybridization term further breaks the gauge symmetry down to a single $U(1)$. The low energy theory is therefore a single QED$_3$ with $N_f=4$ gapless Dirac fermions formed by $f_{a=1,\alpha}$. We will see later that this state is in fact different from the standard $U(1)$ DSL in $S=1/2$ systems in a crucial way: the spin-triplet monopole in this $U(1)_A$ theory carries a trivial lattice momentum ($\vec{\Gamma}$) instead of $\vec{K}$ for the standard $U(1)$ DSL.

\item $U(1)_B$: we can also start from the $U(2)\times U(1)$ ansatz, and add the following alternative terms:
 \bea
 \label{U(1)B}
H_5=&& \Delta_p\sum_{\la\la ij\ra \ra}(-1)^{s_{ij}}(\epsilon_{\alpha\beta}f^{\dagger}_{i,a=2,\alpha}f^{\dagger}_{j,a=3,\beta} +h.c.) \nonumber \\
&&+h\sum_{i}(f^{\dagger}_{i,a=1,\alpha}f_{i,a=2,\alpha}+h.c.).
\eea
This is almost the same with the term for the $U(1)_A$ state, except that $f_{3,\alpha}$ participates in a particle-hole conjugated way ($f_{3,\alpha}\to \epsilon_{\alpha\beta}f^{\dagger}_{3,\beta}$). This term also breaks the gauge symmetry down to a single $U(1)$, and gaps out the Dirac cones formed by the two colors $a=2,3$. The low energy theory is therefore again a QED$_3$ with $N_f=4$. We will see later that this state is identical to the standard $U(1)$ DSL state for $S=1/2$ in terms of universal properties. In particular, the spin-triplet monopoles do carry $\vec{K}$ momentum in the $U(1)_B$ theory. 
\end{enumerate}

As discussed in the main text, all these DSLs have spin-triplet monopoles which contribute to neutron scattering weight. We now analyze the lattice momentum of these triplet monopoles in each theory. Following the discussions in the main text, we conclude that the triplet monopoles in $U(3)$, $U(2)\times\mathbb{Z}_2$,  and $U(1)\times \mathbb{Z}_2\times \mathbb{Z}_2$ states all have momentum $\vec{K}$, which agrees with the experiment. The $U(1)_A$ and $U(1)_B$ phases require more analysis. This can be done by first considering an intermediate state with $U(1)^{(1)}\times U(1)^{(2)}\times U(1)^{(3)}$ gauge field, where the gauge field $A^{(a)}$ of each $U(1)^{(a)}$ only couples to $f_a$. This is simply three copies of QED$_3$, each with triplet monopoles at momentum $\vec{K}$. Now $U(1)_A$ and $U(1)_B$ phases are obtained by Higgsing the $U(1)^{\otimes 3}$ gauge symmetry down to a single $U(1)$, with gauge field $A$. The difference is that for $U(1)_A$ we have $A=A^{(1)}=A^{(2)}=A^{(3)}$ while for $U(1)_B$ we have $A=A^{(1)}=A^{(2)}=-A^{(3)}$. For the monopole operators this means that the monopole operator in $U(1)_A$ phase transforms under various symmetries as  $\mathcal{M}_1\mathcal{M}_2\mathcal{M}_3$, where $\mathcal{M}_a$ is the monopole of $A^{(a)}$; while in $U(1)_B$ phase it transforms as $\mathcal{M}_1\mathcal{M}_2\mathcal{M}_3^{\dagger}$.  Therefore the monopole lattice momentum is $\vec{K}+\vec{K}+\vec{K}=0$ for $U(1)_A$, and $\vec{K}+\vec{K}-\vec{K}=\vec{K}$ for $U(1)_B$. 

We therefore conclude that $U(3)$, $U(2)\times \mathbb{Z}_2$, $U(1)\times \mathbb{Z}_2\times \mathbb{Z}_2$ and $U(1)_B$ phases all contain spin triplet monopoles at $\vec{K}$ that can contribute to neutron scattering spectral weight, consistent with experimental observation\cite{Liuetal20}. We can further demand that the DSL should have a direct, continuous transition to the $120^{\circ}$ magnetic order. This rules out $U(2)\times \mathbb{Z}_2$ and $U(1)\times \mathbb{Z}_2\times \mathbb{Z}_2$ states, since a Gross-Neveu type of transition will not affect the gapped $\mathbb{Z}_2$ topological orders and the resulting states will not be classical (short-range-entangled) magnetic orders. This leaves the $U(3)$ and $U(1)_B$ phases to be the main contenders.

In fact to produce the same low energy theory as the $U(1)_B$ state, we do not even need to have the on-site hybridization term (the $h$-term in Eq.~\eqref{U(1)B}). This will lead to a $U(1)\times U(1)$ gauge symmetry, but the second $U(1)$ is a pure gauge theory without gapless matter, and will confine due to monopole proliferation. From our previous monopole analysis, the monopole of the second $U(1)$ carries lattice momentum $\vec{K}-\vec{K}=0$, so the confinement will not break any physical symmetry. This leaves behind a single $U(1)$ gauge theory with $N_f=4$ gapless Dirac fermions. This is the candidate $U(1)$ DSL for $S=3/2$ discussed in the main text.

Finally we comment on some other mean field ansatz that naively also give DSL-like behaviors. One can start from the $U(3)$ ansatz and make the hopping amplitute $t$ slightly different for each flavor: $t_1\neq t_2\neq t_3$. This breaks the gauge symmetry down to $U(1)\times U(1)\times U(1)$ so the low energy theory becomes three copies of the QED$_3$, each with $N_f=4$ Dirac fermions. This theory, however, is unstable: they physical symmetries allow a monopole-monopole coupling of form $\mathcal{M}_1^{\dagger}\mathcal{M}_2$. The scaling dimension of each monopole is estimated in Monte Carlo simulation\cite{KarthikMonopole} to be $\sim 1.1$ (which is only slightly larger than the large-$N_f$ estimation\cite{DyerMonopoleTaxonomy} $\sim1$). This monopole-monopole coupling is therefore relevant and leads to instability. Since the monopole scaling dimension is expected to decrease as the color number increases, the same instability will  happen if we make $t_1=t_2\neq t_3$, which results in a $U(2)\times U(1)$ QCD theory with monopole-monopole coupling. It is interesting to ask what is the effect of this relevant monopole-monopole coupling. One answer is that it may drive the system through a spontaneous chiral symmetry breaking followed by confinement. A more interesting possibility is that it may drive the system back to the $U(3)$ QCD theory. One motivation for considering this scenario is that a strong monopole-monopole coupling will essentially identify different monopoles and results in a theory with only one type of monopole, but this is exactly what happens in the $U(3)$ QCD theory.

\end{document}